\documentclass{aa501}
\begin{document}

\title{
The quiescent Hubble flow, local dark
energy tests, and pairwise velocity dispersion in a $\Omega = 1$ universe}
\subtitle{}
\author{
P. Teerikorpi \inst{1},
A.D. Chernin \inst{1,2,3},
and Yu.V. Baryshev \inst{4,5}}

\institute{Tuorla Observatory,
University of Turku, FIN-21500 Piikki\"o, Finland
\and
Division of Astronomy, University of Oulu, FIN-90014, Finland
\and
Sternberg Astronomical Institute, Moscow University, 119899, Moscow, Russia
\and
Institute of Astronomy, St.Petersburg State University,
Staryj Peterhoff, 198504, St.Petersburg, Russia
\and
Isaac Newton Institute of Chile, Saint-Petersburg Branch, Russia
}

\date{Received / Accepted}

\abstract{We review the increasing evidence for the cosmological relevance
of the cold local Hubble flow. New observations, N-body
simulations and other theoretical arguments
are discussed, supporting our previous suggestion that the
cosmological vacuum or uniform dark energy can have locally observable
consequencies, especially a lower velocity scatter in DE dominated
regions.
The apparent contradiction between the slight dependence
of the growth factor on $\Omega_{\Lambda}$ and the significant influence
of dark energy in realistic N-body calculations is clarified.
An interesting new result is
that in the standard $\Lambda$ cosmology, gravitation
dominates
around a typical matter fluctuation up to about the correlation
length $r_0$, and we tentatively link this with the high pairwise velocity
dispersion
on scales up to several  Mpc, as measured from galaxy redshift-space
correlations.
Locally, the smooth Hubble flow on similar scales is consistent with
N-body simulations including $\Omega_{\Lambda} \approx 0.7$ and a low
density contrast in the Local Volume, which make it generally
vacuum-dominated beyond $1 - 2$ Mpc from galaxies and groups.
We introduce a useful way to view the Hubble flow in terms of
"zero gravity" spheres
around galaxies: e.g., a set of non-intersecting spheres, observed to be
expanding, actually participates in accelerating expansion.
The observed insensitiveness of the local velocity dispersion to
galaxy mass is explained as an effect of the vacuum, too.
}
\titlerunning{The quiescent Hubble flow and local dark energy tests}

\maketitle

\section{Introduction}

The puzzle of the smooth local Hubble flow was recognized by Sandage et al.
(\cite{sandage72}) and further emphasized by Sandage (\cite{sandage99})
and Thim et al. (\cite{thim03}). From N-body simulations Governato
et al. (\cite{governato97}) predicted a high velocity dispersion
for the Local Group (LG) environment in the case of zero-$\Lambda$
universes.
The low velocity scatter was clearly seen in the data analyzed by 
Ekholm et al. (\cite{ekholm01}) and Karachentsev \& Makarov (\cite{kara01}).
On the theoretical side, Chernin (\cite{chernin01}) argued that
the smooth vacuum density, starting to dominate over the matter
not far from the Local Group (LG), is the dynamical reason for the coldness
of the local Hubble flow. We generalized this novel explanation to
include the time variable dark energy (DE) density (Baryshev et al.
\cite{baryshev01}). 

In recent years new observations and theoretical studies have
only enhanced the importance of the  dynamical properties
of the Local Group environment. New measurements of distances in
the Local Volume
have been made. New N-body simulations relevant to the local Hubble flow
have been performed and a better understanding has been gained on the effect
of the cosmological vacuum and DE on structure formation and peculiar
velocities. Redshift-space correlation studies of deep
galaxies surveys (2dF, SDSS) have given statistical results on the velocity
dispersion on scales comparable to the Local Volume.
Here we review these works and discuss the problem further.

\section{New observations of the local Hubble flow}

After our previous studies of the local Hubble flow (Ekholm et al.
 \cite{ekholm01}; Chernin \cite{chernin01}; Baryshev et al.
\cite{baryshev01}) new observations and analyses have appeared
mostly giving support to the picture of a cold flow in the distance range
$r \la 5 h_{100}^{-1}$ Mpc, which is also
an interesting region in view of the statistical results on
the pairwise velocity dispersion from deep galaxy surveys. 

\subsection{Individual galaxies}

Davidge \& van den Bergh (\cite{davidge01})
measured the distance of the nearby elliptical galaxy Maffei 1 from the
asymptotic
giant branch tip at near-infrared and obtained the result
 $\mu = 28.2 \pm 0.3$ or
$r = 4.4 (+0.6,-0.5)$ Mpc. With a local Hubble constant
 $\approx 60$ kms$^{-1}/$Mpc,
one predicts a recession velocity of about 264 km/s, which well agrees
 with the observed $279 \pm 25$ km/s.

Gieren et al. (\cite{gieren04}) have measured the distance to the nearby
(2 Mpc) galaxy NGC 300; It has a previous HST measurement ($\mu = 26.50$),
but with over one hundred new Cepheids detected from the ground, the
distance could be determined with better accuracy, yielding $\mu
= 26.43 \pm 0.04$ (1.93 Mpc). This galaxy was listed by Teerikorpi \& Paturel
(\cite{teerikorpi02}) as having an unbiased Cepheid distance (c.f. sect.
2.3), and the new measurement appears to confirm that. Its
radial velocity $112$ km/s well corresponds to the predicted
$\approx 116$ km/s.  

Thim et al. (\cite{thim03}) have measured the Cepheid distance to the spiral
galaxy M83 (NGC5236) using the Antu 8.2m telescope of the ESO VLT. From twelve
Cepheids they derived for its dereddened distance modulus the value
$28.25 \pm 0.15$, or a distance
of $4.5 \pm 0.3$ Mpc. This distance was consistent with other available
distances for the group containing M83. The mean recession velocity of
$249 \pm 42$ km/s is again in
agreement with the prediction ($\approx 270$ km/s) if $h_{100} \approx 0.6$.

Rekola et al. (\cite{rekola04}) derived a Cepheid distance to the spiral galaxy
IC342 in the IC342/Maffei group. Their result, $3.8 \pm 0.4$ Mpc, predicts
$V_c \approx 228$ km/s, in comparison with the observed
$V_{\mathrm{LG}} = 230$ km/s.

Rekola et al. (\cite{rekola05}) measured the distance to NGC253 by the planetary nebulae
luminosity function method and in combination with other methods derived
a distance $3.6 \pm 0.2$ Mpc. This predicts $V_c \approx 216$ km/s,
in comparison with the measured $V_{\mathrm{LG}} = 234$ km/s.

\subsection{Karachentsev's TRGB programme}

As one result of the formidable effort to measure distances to as many
as possible
Local Volume galaxies using the luminosity of the tip of the red giant branch
in their programme with the HST and ground telescopes, Karachentsev and his
collaborators  have confirmed their earlier small values
for the velocity dispersion: Karachentsev et al. (\cite{karachentsev02a},
\cite{karachentsev02b}) using the M81 group and
the Centaurus A group, Karachentsev et al. (\cite{karachentsev02c}) in a study
of the very local Huble flow, Karachentsev et al. (\cite{karachentsev03a})
using the Canes Venatici cloud, and Karachentsev et al. (\cite{karachentsev03b})
in a study of local galaxy flows within 5 Mpc.
Karachentsev et al. (\cite{karachentsev03c}) showed that the centroids
of eight nearby galaxy groups have a scatter of about 30 km/s around
the Hubble relation.
 
\subsection{The Cepheid bias and other studies}

The Cepheid stars, generally regarded as the best primary distance indicators,
also contribute to the scatter in the Hubble relation. Teerikorpi \& Paturel
(\cite{teerikorpi02}) and Paturel \& Teerikorpi (\cite{paturel04}) have
presented evidence for a selection bias
in the Cepheid method, which varies from galaxy-to-galaxy,
on average making the measured distances too small. Its influence on
the value of  the Hubble constant and on the local Hubble
diagram has been studied in Paturel \& Teerikorpi ({\cite{paturel05}).

When one considers the whole sample of Cepheid host galaxies up to
the Virgo and Fornax clusters, then the
first order corrections to the bias reduce the dispersion around
the linear Hubble law from 120 to 84 km/s. When one looks at the nearby volume
 with $V_c < 300$ km/s (or $r \la 5$ Mpc, then
the dispersion descends from 35 km/s to 31 km/s. It seems that the large
scatter in the local Cepheid Hubble diagram, which puzzled Freedman et al.
(\cite{freedman01}), was partly due to the bias varying from galaxy-to-galaxy. 

Whiting (\cite{whiting03}) used a sample of local
galaxies with distances derived from various methods and sources. He derives
a velocity dispersion of about 100 km/s and suspects that the smaller
values derived for example by Ekholm et al. (\cite{ekholm01}) are due to
small number statistics. However, the evidence makes it difficult
to reject in this way the reality of a still colder, quite local Hubble flow.

Even a dispersion of around 100 km/s would be interestingly low cosmologically,
as will be discussed below.  We regard with great interest
the result by Whiting (\cite{whiting03}) that the dispersion does not
depend on the mass of a galaxy, but is the same for
giants and dwarfs. Already noticed by Karachentsev \&
Makarov (\cite{karachentsev96}), this phenomenon  certainly deserves
to be studied with larger and more homogeneous samples.

Macci\`{o} et al. (\cite{maccio04}) used a sample of 28 galaxies within about 10 Mpc
(11 with Cepheid-based distances, 17 early types with SBF distances).
From the Hubble diagram they derived the velocity dispersion in spheres of
different sizes around the LG and found that $\sigma_v$ varies from
52 km/s ($r \leq 3$ Mpc) to 135 km/s ($r \leq 10$ Mpc). At small distances,
where the distance indicators have the best accuracy, the
result is in fair agreement with previous works. At larger distances it may
agree with Whiting (\cite{whiting03}).

One may ask if the apparent increase of the dispersion towards larger
distances, clearly seen in Fig.1 of
Macci\`{o} et al. (\cite{maccio04}), could be due to some unaccounted-for
factors.
As the sample extends half-way to the Virgo cluster, some scatter must arise
from the systematic differential infall to Virgo, which would shift some of
the more distant galaxies upwards in the Hubble diagram. However,
such a behaviour is in principle included when one compares observations
and realistic simulations. 
Another effect could be due to the distance indicator. If the derived
distance modulus $\mu$ is affected by a Gaussian error with dispersion
$\sigma_{\mu}$, the inferred velocity dispersion would increase with distance.
In the extreme case, if all dispersion were due to $\sigma_{\mu}$, then
the apparent velocity dispersion $\sigma_v$
would increase linearly with the distance, if the Hubble law is linear.
We would like to know how strong this effect can be and we give
a formula connecting the apparent velocity dispersion $\sigma_v$ with
the rms error $\sigma_{\mu} = \sigma$ in the distance modulus at a fixed
derived distance $R_{\mathrm{der}}$ (Appendix A):
\begin{equation}
\sigma_{\mathrm{v}} 
         \approx H R_{\mathrm{der}}[1 - e^{-0.954\sigma^2}(2e^{-1.70\sigma^2}
         + e^{-2.65\sigma^2})]^{1/2}
\end{equation}
As a more realistic example, let us assume that $\sigma_{\mu} = 0.15$
below 3 Mpc and 0.3 above 6 Mpc. Then at small distances the true velocity
dispersion would be still about 50 km/s. If it
is the same at 10 Mpc (where $V_c \approx 600$ km/s),
then the apparent dispersion would be about 100 km/s.

Hence we regard still uncertain how large the local increase in
$\sigma_v$ with distance actually is. As Macci\`{o} et al. (\cite{maccio04})
demonstrate, determination of this behaviour provides a local cosmological test
when considered together with cosmological N-body simulations, which do
predict such a trend in LG-type environments.

\section{The pair-wise velocity dispersion on Mpc-scales from redshift-space
 correlations}

Recent redshift-space correlation analyses of the SDSS (Hawkins et al.
\cite{hawkins03}) and 2dF (Zehavi et al. \cite{zehavi02}) galaxy surveys
have found high pair-wise peculiar velocity dispersions of 500 - 600 km/s.
The method used in those analyses was applied in the classical work
by Davis \& Peebles (\cite{davis83}) to the small CfA survey.
Such velocity dispersions, together with the cosmic virial theorem, give a
value of about 0.3 for the mass density parameter.

Such a large velocity scatter measured within the scale of
$ \la 10h_{100}^{-1}$ Mpc implies that the typical situation in the galaxy
universe is that around random
galaxies there is practically no detectable Hubble law on such scales.
This undoubtly reflects the fact that most galaxies are members of systems
from groups to superclusters.

The high pair-wise velocity dispersion on scales of the Local Volume or less
is another way to see the problem of the local smooth Hubble flow.
In some sense our local environment, where Edwin Hubble was able to find
his law, is different from the general picture that
emerges from the statistical analysis of redshift-space correlations.
This difference could explain the contrast in velocity dispersions.

\section{$\Lambda$ and the growth rate  of density fluctuation}

Chernin (\cite{chernin01})  suggested that a solution to the problem
of the cold local Hubble flow may be found  in the coincidence that
the antigravity of the cosmological vacuum or dark energy
starts to dominate over the gravity of lumpy matter at the
distance (about 1.5 Mpc) where the Hubble flow emerges.
Thus the relatively low density contrast in the Local Volume
allows vast vacuum-dominated regions. This explanation, which
we generalized to dark energy  (Baryshev et al. \cite{baryshev01}),
might be criticized on the basis that the influence of the $\Lambda$ term
on the growth rate in Friedmann models appears to be miniscule.
We discuss this question in this and the next sections.

\subsection{The growth rate and $\Omega_{\Lambda}$}

 Lahav et al. (\cite{lahav91})
derived the following formula for the growth rate at the present
epoch $f(z = 0) \approx \Omega_m^{0.6} + \frac{1}{70}\Omega_{\Lambda}
(1 + \frac{1}{2} \Omega_m)$. This shows that for a fixed matter density
parameter $\Omega_m$, adding the cosmological vacuum into the model
has practically no effect
for the present growth rate, which also determines peculiar velocities
around the growing density fluctuations. In fact,
when the vacuum density is added, the growth factor slightly increases.

Lahav et al. (\cite{lahav91}) see this insensitivity to $\Lambda$ as reflecting
the cosmic vacuum as a uniform background which does not have local force
effects. A galaxy does not ``feel'' the presence of the vacuum.
The cosmological constant influences the behaviour of the global scale
factor and only in this way enters the differential equations for a growing
individual matter density contrast $\delta$.

\subsection{Adding the constraint
 $\Omega_{\mathrm{m}} + \Omega_{\Lambda} =1$}

On the other hand, we have the condition, from
the fluctuations of the CBR, that the universe has a flat spatial geometry,
so we are constrained to consider the situation $\Omega =\Omega_m +
\Omega_{\Lambda} =1$. Then the present growth
rate will depend significantly on the fraction of the vacuum
$\Omega_{\Lambda}$ in the model. This is clearly seen from the above formula,
and in Fig. 1 of Axenides \& Perivolaropuolos (\cite{axenides02})
for the growth factor of density fluctuations.
The corresponding behaviour of peculiar velocities is seen in Fig. 2 of
Peebles (\cite{peebles84}), in Fig. 8 of Carroll et al. (\cite{carroll92})
and in Fig. 2 of Axenides \& Perivolaropuolos (\cite{axenides02}).

For example,
compared with the case $\Omega_{\Lambda} = 0.0$, peculiar velocities
as calculated from the growth rate
are a factor of 2 smaller when $\Omega_{\Lambda} = 0.7$, while in an extreme
case, a factor of 15 smaller when $\Omega_{\Lambda} = 0.99$.
This does not contradict what was
said above, because now changing the vacuum density is accompanied by
a change of the matter density parameter. 

\section{Dark energy and decay of peculiar velocities}

Another and still more important aspect of the effect of the vacuum follows
from the result that
in the regions of the universe where dark energy dominates new
structures do not condense and linear perturbations of density and peculiar
velocities decay (Chernin \cite{chernin01}; Baryshev et al. \cite{baryshev01};
Chernin et al. \cite{chernin03a},b).

This effect was considered for vacuum-dominated regions by Chernin et al.
(\cite{chernin03a},b) using the method of stability analysis first
suggested by Zeldovich (\cite{zeldovich65}) for Lifshitz-type
perturbations in an expanding universe with $\Lambda = 0$. It was
shown that the inclusion of vacuum can radically change the situation,
so that only decreasing or frozen density perturbations are possible in a
vacuum-dominated region.

It is important that  velocity perturbations, or peculiar velocities,
can only decrease in vacuum-dominated regions, where
the vacuum acts as an effective cooling agent.
In order to see this clearly in a simple situation, we refer
the reader to  eqs. \ref{firstint},\ref{hubblec} in sect. 7.3.
There, eq. \ref{firstint} describes the relative velocity of two masses,
with the asymptotic ($D \rightarrow \infty$)  velocity $V = H_{\mathrm{V}} D
 = D/A_{\mathrm{V}}$
corresponding to Hubble's
linear velocity-distance relation. Deviations from this regular motion are
characterized by  a radial peculiar velocity $v$ which is the difference
between $\dot D$ and $V$. In the simplest case of the parabolic motion,
$E = 0$, eq. \ref{firstint} gives that the peculiar velocity $v$ behaves as:  
\begin{equation}
v = \dot D - D/A_{\mathrm{V}} \propto D^{-2}.
\end{equation}  
This result shows that in vacuum-dominated
regions the vacuum cooling is even more effective than
the usual adiabatic cooling ($v \propto a^{-1}$). 

The end of new structure formation is an effect of the vacuum.
This epoch is usually put at around $z_{\Lambda} \approx 0.7$, when
\emph{on average}
vacuum starts to dominate over gravitating matter in the homogeneous
Friedman universe ($1 + z_{\Lambda} = (2 \Omega_{\Lambda} /
 \Omega_{\mathrm{m}})^{1/3}$). But in the real universe this moment was
not the same everywhere, as the matter density varies within a large
range. That is why it is useful to adopt the concept of a local
"zero-gravity" (ZG) sphere, whose radius can be calculated if one knows
the mass distribution and the vacuum or DE density.
Locally, beyond such a sphere the vacuum dominates and structure
formation has stopped some time ago (the epoch depends on the local matter
density).

\section{N--body simulations of the local environment as cosmological test}

The fluctuation growth rate analysis of peculiar velocities
is concerned with the question of how high velocities are required to
maintain a structure growth (the continuity equation tells that one must
transfer particles to build up the structure). Such an analysis follows the
behaviour of a single growing fluctuation and how the growth is slowed down.
We emphasize that it does not consider the cooling of velocities
in those vacuum-dominated regions where structure formation has stopped.
The net effect of all these processes becomes apparent only in realistic
N-body simulations, such as performed by Klypin et al.
(\cite{klypin03}) and Macci\`{o} et al. (\cite{maccio04}).
Governato et al. (\cite{governato97}) made important simulations
of the local environment within models where $\Omega_{\Lambda} = 0$ and
arrived at high velocity dispersions, e.g. the flat CDM model gave
300 km/s $< \sigma_{\mathrm{v}} < 700$ km/s.

Klypin et al. (\cite{klypin03}) simulated the evolution of a region which was
similar to our environment within 100 Mpc from the LG. They used
the flat $\Lambda$CDM model with $\Omega_{\Lambda} = 0.3$ and obtained
a rather low velocity dispersion of about 60 km/s in the local Hubble flow.

Macci\`{o} et al. (\cite{maccio04}) performed N-body simulations of flat
DE-dominated
universes and showed that galaxies around systems similar to the LG have
low peculiar velocities. They also showed that replacing the cosmological
constant ($w=-1$)
with a DE model with $w=-0.6$, peculiar velocities are still reduced by
about 15 percent (their Fig.4). This is roughly as expected on the basis of
our simple analytic calculation in sect. 5.4 in Baryshev et al.
(\cite{baryshev01}),
where it was argued that the longer adiabatic cooling in the quintessence model
with $w = -2/3$ leads to lower peculiar velocities. One may suppose that
a still lower velocity dispersion might have been found for our third example, 
the coherently evolving model with $w=-2/3$, where the cooling time is still
longer. This model is of special interest
as it produces a Hubble relation that is close to the standard $\Lambda$-model
relation fitting the SNIa observations for $z \la 1.5$ (Teerikorpi et al.
\cite{teerikorpi04a}).

Macci\`{o} et al. (\cite{maccio04}) concluded that two facts are essential
for one to reproduce
the low local velocity dispersion: 1) a correct cosmology, requiring the
$\Lambda$ or DE component, and 2) a correct environmental density
contrast, such
as around the LG, which appears to be rather small (0.2--0.6). If the density
contrast is large, then high peculiar velocities are expected even in DE
cosmologies. This is qualitatively understandable because then the ZG surface
extends to a greater distance and the present gravity-dominated region
is larger.

We note that in the Macci\`{o} et al. simulations the mean value
of the local Hubble constant within a few Mpc is close to the global
one ($70$ kms$^{-1}$/Mpc), with a majority of individual simulated
``Local Volumes'' having values between 60 and 80 (Macci\`{o}, private
communication).

A different kind of local Hubble flow calculation was made by Chernin et al.
(\cite{chernin04}). They traced the trajectories of local galaxies
back to the epoch of the formation of the LG and found initial conditions
that were very different from those that would directly lead to
the linear Hubble flow. With simulations they identified the vacuum as
the agent that introduces the subsequent regularity in the nearby flow.

We conclude that the concepts of gravity- and vacuum-domination as well as
the size of gravity-dominated region are useful tools when one
studies structure formation and the evolution of peculiar velocities.

\section{Characteristics of the zero-gravity surface}

The dynamics  of a spherically symmetric dust matter cloud
with density $\mathrm{\rho_m (r)}$
on the homogeneous DE background
is described by the Einstein's field equations
(Chernin 2001; Baryshev et al. 2001), 
giving the following exact equation of motion:
\begin{equation}
\mathrm{\ddot{r} = - G M_{eff}/r^2; \;\;\; M_{eff} = M_m (r) + M_{DE}(r)}\;,
\label{accel}
\end{equation}
Here $M_{\mathrm{m}}(r) =4\pi \int_0^r \rho_\mathrm{m}(r) r^2 dr$
is the dust mass within the sphere of radius $r$,
 $M_\mathrm{DE}(r)$ is the DE mass within the same radius,
given as $M_{\mathrm{DE}} = (4\pi /3)(1+3w)\rho_{\mathrm{DE}}r^3$.

\subsection{The zero-gravity radius}

For the point-mass model there is a distance $r_\mathrm{ZG}$
where $\ddot{r}=0$ and  the DE gravitating mass
equals that of the matter cloud, i.e.\ $M_{\mathrm{eff}} = 0$.  
For the cosmological constant ($w = -1$) this "zero-gravity radius" is
\begin{equation}
r_{\mathrm{ZG}} = (3M/(8\pi \rho_{\Lambda}))^{1/3} \,.
\label{rlambda}
\end{equation}
For a point mass $r_{\mathrm{ZG}}$ remains always constant.
In the standard flat universe with $\Omega_{\Lambda} = 0.7$
a mass $M_{LG} = 2\, 10^{12} M_{\sun}$ has $r_{ZG} = 1.5\,
h_{60}^{-2/3} \Omega_{\Lambda}^{-1/3}$ Mpc.  
Chernin et al. (\cite{chernin04}) calculated the ZG surface around the
Local Group, dominated by the Milky Way and M31 pair, and found that
it is almost spherical and remains nearly unchanged during a 12.5 Gyr
history of the LG. 

The ZG sphere for a point mass $M$ has special significance in
an expanding universe. A light test particle at
$r > r_{\mathrm{ZG}}$ experiences an acceleration outwards. If it
has even a small recession velocity away from $M$, it participates
in an accelerated expansion.

One may also define another interesting sphere with an ``equal energy''
radius $r_{\mathrm{EE}}$. In such a sphere around the mass $M$
the matter and vacuum energies are equal. Its radius is
somewhat larger than the ZG radius: $r_{\mathrm{EE}} = 2^{1/3}r_{\mathrm{ZG}}
\approx 1.26r_{\mathrm{ZG}}$. This radius means for two identical
point masses the same as the ZG radius for a test particle.
Separated by the distance $D =  r_{\mathrm{EE}}$ the two masses
have zero acceleration relative to the centre-of-mass, while for
$D > r_{\mathrm{EE}}$ they experience outward acceleration.

These examples illustrate in simple situations the general result
that in vacuum-dominated expanding regions, perturbations do not grow (sect.5).

\subsection{The Hubble flow of ZG spheres}

In physics four kinds of mass appear: active
gravitational mass, defined above as $M_{\mathrm{eff}}$, passive
gravitational mass $M_{\mathrm{pas}}$, inertial mass $M_{\mathrm{ine}}$,
and the mass responsible for the gravitational
potential (in Newtonian terms) $M_{\mathrm{pot}}$. These masses
are equivalent for zero-pressure non-relativistic matter.
However, for vacuum the masses per unit volume are $\rho_V +
3 P_V = -2\rho_V$, $\rho_V + P_V = 0$, $ \rho_V + P_V = 0$, and $\rho_V$,
respectively. The equivalence principle tells us that passive
and inertial masses are equivalent; they are both zero for vacuum
that does not feel any gravity and is not affected by matter.

It is interesting and useful to  formulate the Hubble flow in terms
of ZG (or EE) spheres. A typical galaxy
together with its massive halo is well contained within such a cell
($r_{\mathrm{ZG}} \approx 1$ Mpc for $5\,10^{11}M_{\sun})$ as are groups
of the LG type, which are scattered in the Local Volume.
\footnote{However, for a large cluster or a supercluster, the ZG spheres
contain a progressively smaller fraction of the volume.}  
Viewed in this way, cosmological
expansion is the relative motion of the set of ZG spheres. To a first
approximation, the space between them contains just vacuum.

A ZG sphere has zero effective gravitating mass, but non-zero
passive (and inertial) gravity mass, so it feels the global gravity field.
The equation that describes (in the centre-of-mass frame) the relative
motion of two ZG spheres (neglecting first all other ones) has
the mathematical structure of Friedmann's cosmological equation:
\begin{equation}\label{eqmotion}
 \ddot D = D/A_V^2 [1 - (r_1^3 + r_2^3)/D^3],
\end{equation}
where $D$ is the distance between the centres of the cells with (constant)
ZG radii $r_{1}$ and $r_2$, and $A_V = (\frac{8 \pi G}{3} \rho_V)^{-1/2}$.
Note that for two equal masses ($r_1 = r_2 = r_{\mathrm{ZG}}$),
the acceleration is positive, if $D > 2^{1/3}r_{\mathrm{ZG}} =
r_{\mathrm{EE}}$, as was already pointed out above.
In general $\ddot D > 0$ if $D > (r_1^3 + r_2^3)^{1/3}$.
For $r_1 \geq r_2$, this
is always valid if $D > 2^{1/3}r_1$, which means that the mass point 2 lies
outside of the EE sphere of the mass 1.

If we add other mass points sparsely enough so that the above condition is
fulfilled (i.e. each EE sphere contains only its own point mass),
the net accelerations relative to the centre-of-mass remain
positive and if the spheres are originally at rest relative to
each other or are recessing, the system will scatter with
accelerating expansion.  

We may consider eq. \ref{eqmotion} for a spherical distribution
(the radius $= D$) of non-intersecting ZG spheres with radia
$r_{\mathrm{ZG}}$ and a light test particle on its surface. Taking
the equivalent ZG sphere in the centre, it has $R_{\mathrm{ZG}}
= Dr_{\mathrm{ZG}} [(4\pi/3)n)^{1/3}]$ where $n$ is the number
density of the ZG spheres. Then one derives for the test particle
$\ddot D = D/A_V^2 [1 - (4\pi/3)nr_{\mathrm{ZG}}^3] > 0$, which reduces
to the familiar Friedmann equation containing the mean matter density
$\rho_{\mathrm{m}}$ and the vacuum density $\rho_\mathrm{V}$.

A useful implication is the following: \emph{If we see
a region where the galaxies and groups are so sparse that their
EE spheres do not contain other objects
and if this region is expanding, one may
conclude that there is accelerating expansion approaching
the global Hubble expansion rate and no further
structure formation.}

\subsection{Towards the insensitivity to galaxy mass}

Considering further the case of two spheres, the first integral of
the equation of motion is
\begin{equation} \label{firstint}
 \dot D^2 = (D/A_V)^2 [1 + 2 (r_1^3 + r_2^3)/D^3] + 2E.
\end{equation}
For parabolic expansion ($E = 0$), 
the local Hubble "constant" squared becomes
\begin{equation} \label{hubblec}
 H^2 = \frac{8 \pi G}{3} \rho_V [1 + 2 (r_1^3 + r_2^3)/D^3],
\end{equation}
If one forgets other ZG spheres, the Hubble expansion rate for these
two galaxies is close to the value $H_V = 1/A_V$ depending on
the vacuum density only. Because $(r/D)^3$ decreases quickly with increasing
considered distance $D$, such a situation may occur in sparsely populated
regions. Looking at large regions containing many galaxies, the Friedmann
equation and the Hubble expansion rate depend on the vacuum and mean matter
densities (the end of sect.7.2). In the standard cosmology,
$(1 + \rho_{\mathrm{m}}/\rho_{\Lambda})^{1/2} = 1.195$, meaning that the
"full" global
Hubble constant is 20 percent larger than the rate due to vacuum only. For
example, if $H_0 = 72$ kms$^{-1}$/Mpc, $H_\mathrm{V} = 60$ kms$^{-1}/$Mpc.

Eq.\ref{hubblec} gives some insight to why the Hubble law is equally
well followed by massive and light galaxies as noted in sect. 2.3
(Karachentsev \& Makarov \cite{karachentsev96}; Whiting \cite{whiting03}).
For a massive halo of $ 4\,10^{12} M_{\sun} $ the distance $r_i$ is about
1.9 Mpc, hence the factor $2(r_1/D)^3$ is quite small already at $D = 5$ Mpc
and still smaller for smaller haloes. For both massive
and light haloes the square root of the bracketed expression
in eq.\ref{hubblec} differs little from unity on a range of spatial scales.
Thus in this simplified model one measures both for large and small galaxies
practically the same Hubble constant that basically depends on the dominating
vacuum density.

Viewing the Local Volume as a sparse set of ZG spheres, on wide areas
between the groups it is vacuum-dominated (Karachentsev at el.
\cite{karachentsev03c}) and all galaxies follow the accelerating expansion
not far from the global Hubble rate. Whiting (\cite{whiting03}) listed
possible explanations for the mass independence in the local dynamics.
His explanation no.2 would need very massive dark objects that
similarly affect different galaxies. In our explanation the massive
dark vacuum is the agent, but in the sense that it causes
accelerated expansion with its rate varying in a narrow range and which
is automatically accompanied by vacuum cooling, as we explained in
sect.5.

\subsection{Vacuum and the bulk motion}

Although there is a rather regular Hubble flow around us in the Local Volume,
this same volume has a bulk motion relative to the cosmic background
raditation. The motion is similar to that of the LG, or
about 630 km/s (Karachentsev et al. \cite{karachentsev03c} give a summary
of bulk motion measurements).
How is it possible for a regular Hubble expansion to exist
on a scale $\sim H_0R$ superposed on a stream with $V \sim H_0R$?
Chernin (\cite{chernin01}) and Chernin et al. (\cite{chernin03b})
pointed out that the bulk motion and the Hubble flow may co-exist,
because vacuum is co-moving with any motion: two frames of reference may move
with respect to each other with any velocity, but the vacuum looks
identical to them and has the same effects. If it regularizes the Hubble
flow in a vacuum-dominated region which is at rest relative
to the CBR, it does the same for a region moving as a whole relative
to the CBR.
Thus the vacuum appears to be relevant for two aspects of the local Hubble
flow: its regularity and identity with the global flow, and its insensitivity
to the simultaneous large-scale motion.

\section{Typical scales of gravity-dominated regions}

It is possible to calculate \emph{typical scales} for gravity-dominated
regions at the present epoch, assuming that
light traces mass and using the results of galaxy correlation analysis.

\subsection{The ZG radius and the correlation length}

In the classical two-point correlation function analysis, one
assumes that one may present the average fluctuation of the mass density
around a given galaxy as
\begin{equation}
\rho (r) = \bar{\rho}_{\mathrm{m}} [1 + (r/r_0)^{-\gamma}]
\end{equation}
where $\bar{\rho}_{\mathrm{m}}$ is the average mass density, $r_0$ is
the so-called correlation length, and $\gamma$ is the correlation exponent.
Integrating over the distance $r$ from 0 to $r$ one obtains a typical mass
$M_{\mathrm{m}}(r)$ within this scale:

\begin{equation}
M_{\mathrm{m}} (r) = \frac{4\pi}{3} r^3 \bar{\rho}_{\mathrm{m}} [1 + \frac{3}{3-\gamma}
 (r/r_0)^{-\gamma}]
\end{equation}

The vacuum with its constant energy density $\rho_{\Lambda}$ starts
dominating relative to this fluctuation at and beyond the distance where
$M_{\Lambda}(r) \geq M_{\mathrm{m}}(r)$, meaning that the radius of
the ZG sphere $r_{\mathrm{ZG}}$ is obtained from

\begin{equation}
 [1 + \frac{3}{3-\gamma} (r_{\mathrm{ZG}}/r_0)^{-\gamma}]
 = 2 \frac{\rho_{\Lambda}}
   {\bar{\rho_{\mathrm{m}}}}
\end{equation}
With the values generally regarded as standard, i.e.
$r_0 \approx 5h^{-1}_{100}$ Mpc and
$\gamma \approx 1.75$ for the correlation function, and
 $\Omega_{\Lambda} = 0.7$, $\Omega_{\mathrm{m}}
= 0.3$, one calculates
\begin{equation}
r_{\mathrm{ZG}} = 0.8 r_0 \approx 4h^{-1}_{100} \,\, \mathrm{Mpc}
\end{equation}
\subsection{Scales of high pair-wise peculiar velocities}

So we see that typically the gravity-dominated region in the standard
$\Lambda$ universe has a radius not far from the observed correlation
length. The corresponding scale is about $2 r_0$, which is also the
region where the recent redshift-space correlation analyses of the SDSS
(Hawkins et al. \cite{hawkins03}) and 2dF (Zehavi et al. \cite{zehavi02})
galaxy surveys have found high pair-wise peculiar velocity dispersions,
$500 - 600$ km/s. 
We make a few comments in terms of gravity- and vacuum-domination:

1) As such galaxies are now in a gravity-dominated
region, they were also in the past. This is because in a spherically symmetric
fluctuation (without shell crossings) a galaxy at $r_{\mathrm{ZG}}$ marks
this distance all the time (the mass inside the radius defined by the galaxy
remains constant, hence $r_{\mathrm{ZG}}$ is constant for a constant vacuum).

2) The gravity-dominated region within $\approx r_0$ presents a matter
density contrast of $ 2\rho_{\Lambda}/\rho_{\mathrm{m}} -1 \approx 3.7$. 
According to Fig.2 in Macci\`{o} et al. (\cite{maccio04}), one would then
expect in a Local Volume sized sphere a velocity dispersion of
$300 - 400$ km/s, corresponding to a pair-wise dispersion of about 500 km/s,
as observed in SDSS and 2dF.

3) There are indications both in SDSS and 2dF that the pair-wise velocity
dispersion
drops beyond about $2r_0$, i.e. there where the deprojected mutual
distances are generally larger than the ZG distance. It is tempting to suggest
that we are here seeing the same effect as in the Local Volume, where
the smooth Hubble
flow starts immediately after the directly calculated zero-gravity surface.

4) Beyond $10h_{100}^{-1}$ Mpc the measured velocity dispersion is $<< HR$
implying a Hubble
flow. As we are now in the vacuum-dominated region, there is accelerating
expansion and no further structure formation.

\section{Conclusions}

We summarize our conclusions:

\begin{itemize}
\item[$\bullet$]{Recent observations on the Local Volume are consistent with
      a low velocity dispersion around the local Hubble flow and almost the
      same  local and global Hubble constants.}
\item[$\bullet$]{New cosmological N-body simulations by Klypin et al.
(\cite{klypin03}) and Macci\`{o} et al.(\cite{maccio04}) give support to our
      hypothesis that the inclusion of the cosmological vacuum or
        smooth dark energy
      produces lower velocity dispersions in vacuum-dominated
      regions.}
\item[$\bullet$]{The concept of vacuum domination is a useful tool for
         characterizing different regions in the galaxy distribution.
 It is helpful in general to view a region of the galaxy
 universe in terms of "zero-gravity" or "equal-energy" spheres surrounding
galaxies and groups.}
\item[$\bullet$]{Within the standard picture of a two-point correlation
      function for
      galaxy distribution, the vacuum-domination generally starts around
      the correlation length $r_0$ in the standard flat $\Lambda$ universe.}
\item[$\bullet$]{Recent statistical results from the SDSS and 2dF surveys on
        the amplitude
       and scale-dependence of pair-wise velocity dispersion are consistent
      both with our view of the importance of vacuum-domination and with
      the low velocity dispersion in the Local Volume.}
\item[$\bullet$]{Locally, we have the advantage that
      the border of zero-gravity can be directly derived from known
      local masses. The smooth Hubble flow observed beyond this distance
       is consistent with the N-body simulations including
 $\Lambda \approx 0.7$ and a low local density contrast.}
\item[$\bullet$]{The similar local Hubble flow for both massive and light
galaxies may naturally reflect the dynamical dominance of the vacuum.}
\end{itemize}

\begin{acknowledgements}
This study has been supported by The Academy of Finland (project
"Fundamental questions of observational cosmology") and by the
foundation Turun Yliopistos\"{a}\"{a}ti\"{o}. We thank the referee for
useful comments and A. Macci\`{o}
for communicating unpublished results on N-body simulations.
\end{acknowledgements}

\appendix

\section{Velocity scatter and distance errors}

Let us assume that a distance indicator has a Gaussian distribution
of the accidental error in the derived distance modulus, with the
dispersion $\sigma_{\mu} = \sigma$. An error $\delta$ in the modulus, when
the true distance is $R$ and the Hubble law is valid, may
be interpreted as a peculiar velocity $v_{\mathrm{pec}}$ at a constant
cosmological velocity $V$:
\begin{equation}
v_{\mathrm{pec}} = H_0 R (1-10^{0.2 \delta})
\end{equation}
As the error $\delta$ has a Gaussian distribution, the
apparent dispersion squared around the Hubble law is obtained as
 $(H_0 R)^2 \langle (1-10^{0.2 \delta})^2 \rangle$, or:

\begin{equation}
\sigma_v^2 = (H_0 R)^2 \frac{1}{\sqrt{2\pi} \sigma_{\mu}}
\int_{-\infty}^{\infty}
(1 - 2 e^{a\delta} + e^{2a\delta})e^{-\delta^2/2\sigma^2} d\delta
\end{equation}
where $a = 0.2 \ln 10 = 0.46$.

However, as we do not know the true cosmological velocity, it is better
to consider
the apparent velocity dispersion at a \emph{fixed derived
distance}. Then one must write for the peculiar velocity:

\begin{equation}
v_{\mathrm{pec}} = H R_{der} (10^{-0.2 \delta}-1)
\end{equation}
When now calculating the average, one must weight the usual Gaussian error
distribution with the distribution $f(\mu_{\mathrm{app}}-\delta) =
f(\mu)$, giving the relative number of true distance moduli feeding
the subsample having the fixed derived distance modulus $\mu_{\mathrm{der}}
\pm \frac{1}{2} d\mu_{\mathrm{der}}$.
For the case of a radial density distribution $\propto R^{-\alpha}$ and
a magnitude-limited sample:
$f(\mu_{\mathrm{app}}-\delta) \propto e^{k\mu}$,
where $k = 0.2(3-\alpha)ln 10$, the integration results in
\begin{equation}\label{appendix}
\sigma_{\mathrm{v}}
 = H R_{\mathrm{der}} [1 -e^{-\frac{k^2}{2}\sigma^2}(2 e ^{-\frac{(k+a)^2}
       {2}\sigma^2}
            + e ^{-\frac{(k+2a)^2}{2}\sigma^2})]^{1/2}
\end{equation}
E.g., for a homogeneous distribution ($\alpha = 0$):
\begin{equation}
\sigma_{\mathrm{v}} 
         \approx H R_{\mathrm{der}}[1 - e^{-0.954\sigma^2}(2e^{-1.70\sigma^2}
         + e^{-2.65\sigma^2})]^{1/2}
\end{equation}

{}

\end{document}